\documentclass[10pt,a4paper,twocolumn,english,aps,manuscript,pre]{revtex4}
\usepackage[T1]{fontenc}
\usepackage[latin9]{inputenc}
\usepackage{amsmath}
\usepackage{graphicx}

\makeatletter

\pdfpageheight\paperheight
\pdfpagewidth\paperwidth

\providecommand{\tabularnewline}{\\}

\@ifundefined{textcolor}{}
{%
 \definecolor{BLACK}{gray}{0}
 \definecolor{WHITE}{gray}{1}
 \definecolor{RED}{rgb}{1,0,0}
 \definecolor{GREEN}{rgb}{0,1,0}
 \definecolor{BLUE}{rgb}{0,0,1}
 \definecolor{CYAN}{cmyk}{1,0,0,0}
 \definecolor{MAGENTA}{cmyk}{0,1,0,0}
 \definecolor{YELLOW}{cmyk}{0,0,1,0}
}

\DeclareMathSizes{2}{2}{2}{2}
\renewcommand\[{\begin{equation}}
\renewcommand\]{\end{equation}}
\usepackage{hyperref}

\makeatother

\usepackage{babel}
\begin{document}

\title{Dynamics of a Classical Particle in a Quasi Periodic Potential}

\author{Yaniv Tenenbaum Katan}

\author{Physics Department, Technion \textendash{} Israel Institute of Technology,
Haifa 32000, Israel}

\author{Tal Kachman}

\author{Physics Department, Technion \textendash{} Israel Institute of Technology,
Haifa 32000, Israel}

\author{Shmuel Fishman}

\author{Physics Department, Technion \textendash{} Israel Institute of Technology,
Haifa 32000, Israel}

\author{Avy Soffer}

\author{Mathematics Department, Rutgers University, New-Brunswick, NJ 08903,
USA}
\begin{abstract}
We study the dynamics of a one-dimensional classical particle in a
space and time dependent potential with randomly chosen parameters.
The focus of this work is a quasi-periodic potential, which only includes
a finite number of Fourier components. The momentum is calculated
analytically for short time within a self-consistent approximation,
under certain conditions. 

We find that the dynamics can be described by a model of a random
walk between the Chirikov resonances, which are resonances between
the particle momentum and the Fourier components of the potential.
We use numerical methods to test these results and to evaluate the
important properties, such as the characteristic hopping time between
the resonances. This work sheds light on the short time dynamics induced
by potentials which are relevant for optics and atom optics.
\end{abstract}
\maketitle

\section{Introduction}

Random potentials have been studied broadly for over a 100 years \cite{lemons1997paullangevintextquoterights,uhlenbeck1930onthe,sturrock1966stochastic,Chirikov1979263,golubovic1991classical,levi-naturephys-2012}.
In particular, random potentials that vary in both time and space
were investigated extensively \cite{levi-naturephys-2012,wilkinson1991adiabatic,krivolapov2012transport,krivolapov2012universality,yukalov2009bose,bourgain1999growthof}.
Despite the great interest that this type of potentials attract, only
partial understanding of the effect of random potentials which fluctuate
in both space and time currently exists. The understanding of quasi-periodic
\cite{bourgain1999growthof,borgonovi_particle_1997} potentials of
this type is even less satisfactory.

In the simplified case in which the potential is time independent,
Anderson localization \cite{PhysRev.109.1492,PhysRevLett.42.673,john1984electromagnetic}
for waves is found \cite{citeulike:1134994,billy2008direct}. In the
more general scenario in which the random potentials also depend on
time, the arguments for Anderson localization break and it is not
generally clear what would be the type of dynamics in these systems.

The study of random potentials which modulate in space and time is
greatly motivated by experimental setups. A common example for such
potentials are optical setups. In optics experiments, the analogy
to the Schrödinger equation is achieved by means of the paraxial approximation
\cite{sulem1999nonlinear} and a slowly varying envelope approximation
\cite{saleh2007fundamentals}. Under these approximations, fluctuations
of the refractive index in the longitudinal direction can be described
by random potentials which depend on time. It follows that a viable
description of light propagation in fluctuating media requires understanding
of the effect of random potentials which depend on both time and space.
In particular, many optical experiments found that intriguing effects
can be generated by quasi-periodic potentials \cite{Quasiperiodic_Optical_Lattices,Yuce2013}.
Thus, understanding the effect of fluctuating potentials which are
quasi-periodic in both space and time is an important goal.

In the present work we consider the classical dynamics in a quasi
periodic potential of the form
\begin{equation}
V\left(x,t\right)=\frac{1}{\sqrt{N}}\sum_{m=-N}^{N}A_{m}e^{-i\left(k_{m}x-\omega_{m}t+\phi_{m}\right)}+cc,\label{eq:V(x,t)-1}
\end{equation}
where $\left\{ A_{m}\right\} $, $\left\{ k_{m}\right\} $ and $\left\{ \omega_{m}\right\} $
are real random numbers. The phases $\left\{ \phi_{m}\right\} $ are
independent random variables, uniformly distributed between $0$ to
$2\pi$. 

This is the natural form of random potentials that are introduced
in optics \cite{levi-naturephys-2012,citeulike:1134994} and in atom
optics \cite{billy2008direct}. In the $N\rightarrow\infty$ limit,
Eq.(\ref{eq:V(x,t)-1}) can be considered as a random potential, while
for finite $N$, it is quasi-periodic. As $N$ increases, the potential
appears more random.

It is believed that for high momentum, classical treatment of such
systems is appropriate for the description of some aspects of their
wave dynamics. This is based on the correspondence principle. Following
this approach, we will focus on classical particles.

The effect of Eq.(\ref{eq:V(x,t)-1}) was recently studied in the
large $N$ limit \cite{krivolapov2012transport,krivolapov2012universality,levi-naturephys-2012}.
It was obtained that the dynamics of this system over large time scales
can be described by a Fokker-Planck equation with momentum dependent
coefficients, yielding anomalous diffusion in phase space. However,
the scenario in which $N$ is not large in Eq.(\ref{eq:V(x,t)-1})
was not studied. Thus, it is not well understood what is the type
of dynamics in such a scenario.

In the present paper, on the other hand, we study the behavior of
Eq.(\ref{eq:V(x,t)-1}) over short time scales, in the small $N$
limit, in which Eq.(\ref{eq:V(x,t)-1}) is quasi-periodic.

As a result of the action of the quasi periodic potential (\ref{eq:V(x,t)-1}),
the phase space is mixed and in some parts the motion is chaotic,
while in other parts it is regular \cite{tabor,book_mechlichtenberglieberman,RevModPhys.64.795}. 

We demonstrate that under certain assumptions, the trajectories of
classical particles in this system are composed of segments in which
momentum oscillates around constant values that are related to the
Chirikov resonances, which are resonances between the momentum and
the Fourier components of the potential \cite{Chirikov1979263,zaslavskiui1972stochastic}.
We calculate the rate of change of the position in these segments
as self consistent approximation and establish that this type of dynamics
can be described by a model of random walk between resonances in phase
space. We then use numerical methods to solve the equations of motion,
to evaluate important characteristic of this system, such as the hopping
time between resonances, and to validate our results. Furthermore,
we demonstrate numerically that our results are applicable to much
stronger potentials then has been previously considered \cite{krivolapov2012transport,levi-naturephys-2012,PhysRevLett.69.1831,bezuglyy2006generalized}. 

The outline of this paper is as follows. In Sec. (\ref{sub:General-Formalism})
the model is defined, and the relevant regime is characterized. In
Sec.(\ref{sub:Short-Time-Scales}), the model is analyzed analytically
over short time scales, and these scales are estimated, while in Sec.(\ref{sub:Finite-Time-Scales})
it is explored for larger time scales. In Sec.(\ref{sub:Comparison-to-Numerics}),
the validity of the approximate results of the previous sections is
tested numerically. The results are then summarized in Sec.(\ref{sec:Summary-and-Discussion}),
and their relevance to the general field is discussed.

\section{The Model \label{sub:General-Formalism}}

We investigate a specific model system and study the dynamics of a
classical particle, described by the Hamiltonian 
\begin{equation}
H\left(x,t\right)=\frac{p^{2}}{2m}+V\left(x,t\right),\label{eq:Full Hamiltonian}
\end{equation}
where $V\left(x,t\right)$ is a one dimensional potential of the form
(\ref{eq:V(x,t)-1}). For convenience, we write Eq. (\ref{eq:V(x,t)-1})
in the form 
\begin{equation}
V\left(x,t\right)=\frac{2A}{\sqrt{N}}\sum_{m=1}^{N}\cos\left(k_{m}x-\omega_{m}t+\phi_{m}\right),\label{eq:V(x,t)}
\end{equation}
 and take the mass in Eq.(\ref{eq:Full Hamiltonian}) to be unity. 

Early work by Chirikov et. al. \cite{Chirikov1979263,chirikov1982diffusion,zaslavskiui1972stochastic}
studied the dynamics of Eq.(\ref{eq:Full Hamiltonian}) in the limit
of extremely small amplitudes. It was predicted that the dynamics
of this system is governed by the structure of the ``Chirikov resonances''
in phase space. These resonances, denoted as $\left\{ P_{m}^{res}\right\} $
are defined through the stationary phase condition,

\begin{equation}
0=\frac{d}{dt}\left(k_{m}x-\omega_{m}t+\phi_{m}\right)=k_{m}P_{m}^{res}-\omega_{m}.\label{eq:stationary phase res requirement}
\end{equation}
Equation (\ref{eq:stationary phase res requirement}) can be simplified
to 
\[
P_{m}^{res}=\frac{\omega_{m}}{k_{m}}.
\]
The resulting equations of motion are

\begin{equation}
\begin{array}{ccl}
\dot{p} & = & -\frac{\partial H}{\partial x}=\frac{2A}{\sqrt{N}}\sum_{m=1}^{N}k_{m}\sin\left(k_{m}\left(x-P_{m}^{res}t\right)+\phi_{m}\right),\\
\dot{x} & = & \frac{\partial H}{\partial p}=p.
\end{array}\label{eq:EOM-1}
\end{equation}

Note that the transformation $x\left(t\right)\rightarrow x\left(t\right)+P_{m}^{res}t$
is a Galilean transformation into the frame of the $m^{th}$ term
in Eq.(\ref{eq:V(x,t)}). We order the resonances by writing 
\[
P_{1}^{res}\ldots\leq P_{m}^{res}\leq P_{m+1}^{res}\leq\ldots P_{N}^{res}.
\]
The phase space distance between two adjacent resonances $P_{m}^{res}$
and $P_{m+1}^{res}$ is defined as $\tilde{\Delta}{}_{m}=P_{m+1}^{res}-P_{m}^{res}$.
The initial conditions are chosen such that the initial momentum is
near a Chirikov resonance $P_{n}^{res},$ and the position is such
that the argument of the cosine with $n=m$ in \eqref{eq:V(x,t)}
is small in a way that is precisely defined before Eq. \eqref{eq:Pstat}.
We will demonstrate that these initial conditions lead to the result
\eqref{eq:26}. 

According to the Chirikov description, when the momentum of the particle
approaches a resonance, it initiates an oscillatory motion around
this resonance. We can obtain this result by assuming that the main
contribution to the force comes from the resonant term and omitting
the remaining Fourier components in Eq.(\ref{eq:V(x,t)}). For an
isolated resonance the system can be reduced to a mathematical pendulum
by a simple Galilean transformation. The momentum is then found to
be 
\begin{equation}
p\left(t\right)=P_{n}^{res}+p_{osc}\left(t\right).\label{eq:x perndulum}
\end{equation}
where $p_{osc}\left(t\right)$ is the oscillatory component. The width
of the resonances, is determined from energy conservation as
\[
\Delta\approx\sqrt{\frac{4A}{\sqrt{N}}\left|\cos\pi-\cos0\right|}=\sqrt{\frac{8A}{\sqrt{N}}},
\]
such that to first order in $\Delta$, the momentum oscillates between
$P_{n}^{res}+\Delta$ and $P_{n}^{res}-\Delta$. $P_{n}^{res}$ therefore
``traps'' the momentum of the particle, in the sense that when $p\left(t\right)$
approaches $P_{n}^{res}$ at a time $t$, it remains close to $P_{n}^{res}$
at later times. 

This description is only valid in the limit of small resonance's widths,
\begin{equation}
\Delta<\tilde{\Delta}{}_{k}\forall k.\label{eq:Isolated Condition}
\end{equation}
In this case, the resonances of Eq. (\ref{eq:V(x,t)}) are separated
in phase space. Thus, when the momentum of a particle is close to
a resonance $P_{n}^{res}$, it is distant from the other resonances. 

Assuming that the effect of the non-resonant terms is negligible as
described by \cite{Chirikov1979263,chirikov1982diffusion,zaslavskiui1972stochastic},
the effect of $P_{n}^{res}$ on the momentum can be expected to be
similar to that of an isolated resonance. In the present work, however,
this only holds for a limited time since the resonances considered
here are not completely isolated. One should remember that in the
studied system ($N$ is finite and small) the potential $V\left(x,t\right)$
oscillates with several frequencies, therefore the Chirikov resonance
picture for our model differs from the one found in some earlier approximations
\cite{chirikov1982diffusion}. 

In the present work, we study the dynamics generated by (\ref{eq:Full Hamiltonian})
for finite and small number of terms in (\ref{eq:V(x,t)}) and for
a weak overlap between the resonances. Concretely, we assume that
$\Delta$ ,the width of the Chirikov resonances, is smaller or of
the order of their separation,
\begin{equation}
\frac{\Delta}{\left\langle \tilde{\Delta}{}_{m}\right\rangle }<1,\label{eq:delta}
\end{equation}
where $\left\langle \tilde{\Delta}{}_{m}\right\rangle $ is the characteristic
phase space distance between adjacent resonances. In such a situation
the overlap between resonances is small and vanishes in the leading
order of the theory \cite{Chirikov1979263,zaslavskiui1972stochastic},
but not exactly. Thus, in the present work the resonances are not
isolated, but overlap weakly.

\section{Short Time Scales\label{sub:Short-Time-Scales}}

We first focus on the dynamics at the time $0<t<t_{hop}$, in which
the momentum remains in the proximity of a resonance $P_{n}^{res}$.
We will demonstrate that at this time interval, the trajectory of
the particle is governed by $P_{n}^{res}$. The dynamics in the other
time intervals is found to be similar.

We write $x\left(t\right)$ as a sum of linear and oscillating components
\begin{equation}
\begin{array}{c}
x\left(t\right)=p_{stat}t+\xi\left(t\right)\end{array}\label{eq:x(t) n res}
\end{equation}
where $\xi$ is the oscillating component of the trajectory.

Note that Eq.(\ref{eq:x(t) n res}) does not uniquely define $p_{stat}$
and $\xi\left(t\right)$. We will impose conditions that ensure a
well defined value of $p_{stat}$ in later parts of this section. 

For the time being, we focus on the case in which $p_{stat}$ is extremely
close to the $n^{th}$ resonance, namely 
\begin{equation}
\varepsilon=\frac{\left|p_{stat}-P_{n}^{res}\right|}{min\left\{ \tilde{\Delta}{}_{n},\tilde{\Delta}{}_{n-1}\right\} }\ll1.\label{eq:Eps}
\end{equation}
Note that $\varepsilon$ depends on the initial conditions and can
in principal receive any value. Yet, we verify by numerical methods
(see Sec. (\ref{sub:Comparison-to-Numerics})) that the inequality
(\ref{eq:Eps}) typically holds (see Fig. \ref{fig:P_delta}(b)) under
the given assumptions.

Since the resonances are well separated, $P_{n}^{res}$ is isolated
from the remaining resonance spectrum. This is consistent with the
condition 
\begin{equation}
\eta=max\left\{ \frac{\Delta}{\tilde{\Delta}{}_{n}},\frac{\Delta}{\tilde{\Delta}{}_{n-1}}\right\} \ll1.\label{eq:delta-1}
\end{equation}
$\eta$ is the inverse of the minimum phase space separation between
$P_{n}^{res}$ and the other resonances. The small parameters of this
problem are $\varepsilon$ and $\eta$, where we assume $\varepsilon\approx\eta\ll1$
. We will calculate the dynamics to the leading order of these parameters. 

We will focus on a small scale, where the short time scale condition
is formulated as 
\begin{equation}
\omega_{n}t\leq\mathcal{O}\left(\frac{1}{\eta}\right).\label{eq:wnt}
\end{equation}
We denote the separation of $p_{stat}$ from the $n^{th}$ resonance
by 
\begin{equation}
\delta p_{n}=p_{stat}-P_{n}^{res}.
\end{equation}
It follows that \textbf{
\begin{equation}
k_{n}\delta p_{n}t\leq\mathcal{O}\left(\varepsilon\right).\label{eq:kn<1}
\end{equation}
}The condition (\ref{eq:kn<1} ) follows directly from (\ref{eq:Eps})
and (\ref{eq:wnt}). To demonstrate this, we use (\ref{eq:Eps}) and
the equality $P_{n}^{res}=\frac{\omega_{n}}{k_{n}}$ to write (\ref{eq:kn<1})
in the form
\[
\frac{\Delta}{P_{n}^{res}}\varepsilon\omega_{n}t\leq\mathcal{O}\left(\varepsilon\right).
\]
Since the resonances are well separated (but not completely), we can
in general expect that $\frac{\Delta}{P_{n}^{res}}\leq\mathcal{O}\left(\eta\right)$
and (\ref{eq:kn<1}) is therefore satisfied.

In this section we focus on short time scales , in which (\ref{eq:kn<1})
holds, while in the following section we obtain a description for
the dynamics over longer time scales. In particular, we will use the
point in time which satisfies (\ref{eq:kn<1}), as an estimate for
the time in which momentum is no longer in the vicinity of the resonance
$P_{n}^{res}$. 

We continue to extract an expression for $p_{stat}$ from the equations
of motion, Eq.(\ref{eq:EOM-1}). With the help of Eq.(\ref{eq:x(t) n res}),
Eq. (\ref{eq:EOM-1}) takes the form
\begin{equation}
\begin{array}{ccl}
\ddot{\xi} & =- & \frac{2A}{\sqrt{N}}\sum_{m}\frac{k_{m}}{\delta p_{m}}\sin\left(k_{m}\left(\delta p_{m}t+\chi_{m}\left(t\right)\right)\right).\end{array}\label{eq:EOM for xosl+xconst}
\end{equation}
where $\chi_{m}\left(t\right)=\xi\left(t\right)+\frac{\phi_{m}}{k_{m}}$.
We can write Eq.(\ref{eq:EOM for xosl+xconst}) in the form 

\begin{equation}
\begin{array}{ccl}
\ddot{\xi} & = & \frac{2A}{\sqrt{N}}\sum_{m}k_{m}\left\{ \sin\left(k_{m}\delta p_{m}t\right)\cos\left(k_{m}\chi_{m}\right)\right.\\
 &  & \left.+\cos\left(k_{m}\delta p_{m}t\right)\sin\left(k_{m}\chi_{m}\right)\right\} .
\end{array}\label{eq:EOM2}
\end{equation}
We then integrate Eq.(\ref{eq:EOM2}) in parts to obtain
\begin{equation}
\begin{array}{ccl}
p_{stat}+\dot{\xi}\left(t\right) & = & \dot{\xi}\left(0\right)+\frac{2A}{\sqrt{N}}\sum_{m}\frac{1}{\delta p_{m}}\left\{ \cos\left(k_{m}\xi\left(0\right)+\phi_{m}\right)\right.\\
 &  & -\cos\left(k_{m}\delta p_{m}t\right)\cos\left(k_{m}\chi_{m}\left(t\right)\right)\\
 &  & +\sin\left(k_{m}\delta p_{m}t\right)\sin\left(k_{m}\chi_{m}\left(t\right)\right)\\
 &  & -k_{m}\int_{0}^{t}\cos\left(k_{m}\delta p_{m}t'\right)\sin\left(k_{m}\chi_{m}\right)\dot{\chi}_{m}dt'\\
 &  & \left.-k_{m}\int_{0}^{t}\sin\left(k_{m}\delta p_{m}t'\right)\cos\left(k_{m}\chi_{m}\right)\dot{\chi}_{m}dt'\right\} .
\end{array}\label{eq:EOM_3_1}
\end{equation}
This can be brought to the form
\begin{equation}
\begin{array}{ccl}
p_{stat}+\dot{\xi} & = & \dot{\xi}\left(0\right)+\frac{2A}{\sqrt{N}}\sum_{m}\frac{1}{\delta p_{m}}\left\{ \cos\left(k_{m}\chi_{m}\right)\right.\\
 &  & -k_{m}\int_{0}^{t}\sin\left(k_{m}\left(\chi_{m}+\delta p_{m}t'\right)\right)\dot{\xi}dt'\\
 &  & \left.-\cos\left(k_{m}\left(\chi_{m}+\delta p_{m}t\right)\right)\right\} .
\end{array}\label{eq:EOM_3}
\end{equation}

Equation (\ref{eq:EOM_3}) is exact for any general decomposition
of the form (\ref{eq:x(t) n res}). The contribution of the non-resonant
terms, $m\neq n$, to the RHS Eq.(\ref{eq:EOM_3}) is of $\mathcal{O}\left(\eta\right)$.
This last statement can be made clear by using the condition $\varepsilon\ll1$
to replace $\frac{2A}{\sqrt{N}\delta p_{m}}$ with $\frac{2A}{\sqrt{N}\Delta_{m}}=\Delta\times\mathcal{O}\left(\eta\right)$
in these terms. 

We now expand the RHS of Eq.(\ref{eq:EOM_3}) in powers of $\varepsilon$.
To first order in $\varepsilon$, this procedure consists of omitting
the non-resonant terms with $m\neq n$ from the RHS of Eq.(\ref{eq:EOM_3}),
which then takes the form 
\begin{equation}
\begin{array}{ccl}
p_{stat}+\dot{\xi} & = & \dot{\xi}\left(0\right)+\frac{2A}{\sqrt{N}}\frac{1}{\delta p_{n}}\left\{ \cos\left(k_{n}\xi\left(0\right)\right)\right.\\
 &  & -k_{n}\int_{0}^{t}\sin\left(k_{n}\left(\xi\left(t'\right)+\delta p_{n}t'\right)\right)\dot{\xi}dt'\\
 &  & \left.-\cos\left(k_{n}\left(\xi\left(t\right)+\delta p_{n}t\right)\right)\right\} +\mathcal{O}\left(\eta\right).
\end{array}\label{eq:EOM_4-1}
\end{equation}

We will assume $k_{n}\xi\left(t\right)=\mathcal{O}\left(\varepsilon\right)$,
$\frac{\dot{\xi}}{\Delta}=\mathcal{O}\left(\varepsilon\right)$, and
proceed to estimate Eq.(\ref{eq:EOM_4-1}) as a self consistent approximation
that will be justified in what follows (subsection \ref{sub:Self-Consistency-of}),
\begin{equation}
\begin{array}{c}
\begin{array}{c}
p_{stat}=-\frac{2A}{\sqrt{N}}\frac{1}{\delta p_{n}}\cos\left(k_{n}\delta p_{n}t\right)+\mathcal{O}\left(\eta\right)\\
=-\frac{2A}{\sqrt{N}}\frac{1}{\delta p_{n}}+\mathcal{O}\left(\varepsilon\right)+\mathcal{O}\left(\eta\right),
\end{array}\end{array}\label{eq:Pstat}
\end{equation}
leading to
\[
p_{stat}\left(P_{n}^{res}-p_{stat}\right)=\frac{2A}{\sqrt{N}}+\mathcal{O}\left(\varepsilon\right)+\mathcal{O}\left(\eta\right).
\]
By solving to leading order, we obtain
\begin{equation}
p_{stat}=P_{n}^{res}-2\frac{A}{\sqrt{N}P_{n}^{res}}+\mathcal{O}\left(\varepsilon\right)+\mathcal{O}\left(\eta\right)\label{eq:26}
\end{equation}
which is a closed expression for $p_{stat}$. Note that under the
self-consistent assumptions $\frac{\dot{\xi}}{\Delta}=\mathcal{O}\left(\varepsilon\right)$
and $k_{n}\chi_{n}\left(t\right)=\mathcal{O}\left(\varepsilon\right)$,
Eq.(\ref{eq:26}) is unique at the short time interval, defined by
(\ref{eq:wnt}).

Another important result is that at the limit $A\rightarrow0$ we
have $p_{stat}\rightarrow P_{n}^{res}$, in agreement with \cite{Chirikov1979263,zaslavskiui1972stochastic}.
A significant notion is that $p_{stat}$ only depends on the details
of the potential, and not on the initial conditions, provided the
initial momentum is sufficiently close to one of the Chirikov resonances.
Explicitly, the assumptions on the initial conditions are that the
momentum is in the vicinity of a resonance $P_{n}^{res}$ and $k_{n}\chi_{n}\left(0\right)$
is smaller of $\mathcal{O}\left(\varepsilon\right)$ ($\chi_{n}$
is defined after Eq. (\ref{eq:EOM for xosl+xconst})).

Let us now examine the case in which momentum varies significantly
from all resonances. We can assume that in this case, $\left\langle \tilde{\Delta}_{n}\right\rangle $
and $\min{}_{m}\left\{ \delta p_{m}\right\} $ are comparable in magnitude,
such that the phase space distance between $p\left(t\right)$ and
the resonances is of the same order as the distance between resonances.
It follows that the terms $\frac{2A}{\sqrt{N}}\frac{1}{\delta p_{m}}$
on the RHS of Eq.($\ref{eq:EOM_3}$) are of $\mathcal{O}\left(\eta\right)$.
Then, to first order in $\eta$, 
\[
p\left(t\right)=p\left(t=0\right)+\mathcal{O}\left(\eta\right).
\]

\subsection{Self Consistency of the Approximation\label{sub:Self-Consistency-of}}

In Eq.(\ref{eq:Pstat}), we assumed that the $\dot{\xi}$ term is
of $\mathcal{O}\left(\varepsilon\right)$. We will justify this treatment
in what follows. We use the fact that Eq.(\ref{eq:Pstat}) is linear
in $\dot{\xi}$, to write $\dot{\xi}=\dot{\xi}_{0}+\dot{\xi}_{1}$,
where the term $\dot{\xi}_{0}$ is related to the effect of $P_{n}^{res}$,
while the term $\dot{\xi}_{1}$ is related to the effect of the remaining
resonances. We demonstrate in a self consistent manner that one can
take $\frac{\dot{\xi}_{0}}{\Delta}=\mathcal{O}\left(\varepsilon\right)$
and $\frac{\dot{\xi}_{1}}{\Delta}=\mathcal{O}\left(\eta\right)$. 

We first assume $\frac{\dot{\xi}_{1}}{\Delta}=\mathcal{O}\left(\eta\right)$.
Then, the equation of motion for $\dot{\xi}_{0}$ is
\[
\begin{array}{c}
\dot{\xi}_{0}=-\frac{2A}{\sqrt{N}}\frac{1}{\delta p_{n}}\left(1-\cos\left(k_{n}\left(\delta p_{n}t+\xi_{0}\right)\right)\right)\\
-k_{n}\int\sin\left(k_{n}\left(\delta p_{n}t+\xi_{0}\right)\right)\dot{\xi}_{0}dt.
\end{array}
\]
Assuming $\frac{\dot{\xi}_{0}}{\Delta}$ is bounded by $\mathcal{O}\left(\varepsilon\right)$
on the RHS, one finds that $\frac{\dot{\xi}_{0}}{\Delta}$ is bounded
by $\mathcal{O}\left(\varepsilon\right)$ in the LHS, provided $k_{n}\left(\delta p_{n}t+\xi_{0}\right)=\mathcal{O\left(\varepsilon\right)}$. 

We proceed to obtain that $\frac{\dot{\xi}_{1}}{\Delta}=\mathcal{O}\left(\eta\right)$.
In a similar manner, the equation of motion for $\dot{\xi}_{1}$ is
given by

\begin{equation}
\begin{array}{c}
\begin{array}{ccl}
\left|\dot{\xi}_{1}\right| & = & \left|\sum_{m\neq n}\frac{2A}{\sqrt{N}}\frac{1}{\delta p_{m}}\left\{ \cos\left(k_{m}\left(\chi_{m}+\delta p_{m}t\right)\right)\right.\right.\\
 &  & \left.\left.+k_{m}\int\sin\left(k_{m}\left(\chi_{m}+\delta p_{m}t\right)\right)\dot{\xi}dt\right\} \right|\\
 & \leq & \eta\left|\sum_{m\neq n}\left\{ \cos\left(k_{m}\left(\chi_{m}+\delta p_{m}t\right)\right)\right.\right.\\
 &  & \left.\left.+k_{m}\int\sin\left(k_{m}\left(\chi_{m}+\delta p_{m}t\right)\right)\dot{\xi}dt+\mathcal{O}\left(\varepsilon\right)\right|\right\} .
\end{array}\end{array}\label{eq:EOM_5}
\end{equation}
Using $\frac{\dot{\xi}_{0}}{\Delta}=\mathcal{O}\left(\varepsilon\right)$,
Eq.(\ref{eq:EOM_5}) becomes 
\[
\begin{array}{c}
\begin{array}{ccl}
\left|\dot{\xi}_{1}\right| & \leq & N\eta\times\max_{j}|\cos\left(k_{m}\left(\chi_{m}+\delta p_{m}t\right)\right)\\
 &  & +k_{m}\int\sin\left(k_{m}\left(\chi_{m}+\delta p_{m}t\right)\right)\dot{\xi}_{1}dt+\mathcal{O}\left(\varepsilon\right)|.
\end{array}\end{array}
\]
Provided $N$ is finite and small, and that $\frac{\dot{\xi}_{1}}{\Delta}$
is bounded by $\mathcal{O}\left(\eta\right)$ on the RHS, it is bounded
by $\mathcal{O}\left(\eta\right)$ in the LHS, in agreement with the
assumption (\ref{eq:Pstat}). We examine the distribution of $\frac{\dot{\xi}}{\Delta}$
by numerical means in Sec. \ref{sub:Comparison-to-Numerics}. We find
a that indeed, for the vast majority of the segments $\frac{\dot{\xi}}{\Delta}\ll1$,
as expected.

We can use similar methods to obtain that $k_{n}\xi=\mathcal{O\left(\varepsilon\right)}$,
provided $\frac{\dot{\xi}_{0}}{\Delta}=\mathcal{O\left(\varepsilon\right)}$
.

\section{Finite Time Scales\label{sub:Finite-Time-Scales}}

The dynamics described so far is only valid for a short time interval,
in which (\ref{eq:wnt}) is satisfied; Since the resonances are not
completely isolated, the momentum will eventually approach a different
resonance $P_{n}^{res}\neq P_{m}^{res}$ at a time point denoted by
$t_{hop}$, leading to the breaking of Eq.(\ref{eq:26}). Nevertheless,
the above description can be generalized to longer time scales $t>t_{hop}$,
as explained in what follows.

Assuming that at a short time interval after $t_{hop}$, $t_{hop}<t<t_{hop}+\delta t$,
where $k_{n}\delta p_{n}\delta t=\mathcal{O}\left(\varepsilon\right)$,
the phase space distance between the momentum of the particle and
the new resonance $P_{m}^{res}$ is of $\mathcal{O}\left(1\right)$,
$P_{m}^{res}$ governs the motion of the particle at this time interval.
We can then apply the analysis presented in the previous section (Sec.
\ref{sub:Short-Time-Scales}) with $\varepsilon_{m}=\frac{\left|p_{stat}-P_{m}^{res}\right|}{min\left\{ \tilde{\Delta}{}_{m},\tilde{\Delta}{}_{m-1}\right\} }$
in the role of the small parameter $\varepsilon$. Thus, at the time
interval $t_{hop}<t<t_{hop}+\delta t$, $x\left(t\right)$ can be
described by Eq.(\ref{eq:x(t) n res}), where $p_{stat}$ is now given
by (\ref{eq:26}) with $P_{m}^{res}$ in the role of $P_{n}^{res}$.
Within the Chirikov picture, this is analogous to a ``hop'' between
two resonances in phase space.

By applying a similar treatment, we can deduce that over longer time
scales, the momentum hops between the phase space resonances, such
that between two consecutive hops, it remains close to a single resonance.
Note that the hopping process is governed by the oscillatory component
$\xi$, which depends on the non resonant terms in Eq.(\ref{eq:V(x,t)}).
The hopping process is therefore stochastic, such that the hopping
between the phase space resonances appears random. Thus, the dynamics
in phase space is analogous to a random walk between resonances.

According to this picture, the trajectory of the particle in real
space is a composition of linear segments, where at each segment $x\left(t\right)$
weakly oscillates around $p_{stat}t$. $p_{stat}$ is given by Eq.(\ref{eq:26})
and is therefore determined by the magnitude of $A$ and of the resonance
$P_{n}^{res}$. The magnitude of the oscillations is determined by
the phase space distance between the resonances and by the magnitude
of $\varepsilon_{m}$. 

Following the assumption that the resonances are well (but not completely)
separated, we can expect that the rate of hopping between resonances
is finite. Furthermore, the hopping time, $t_{hop}$, can be expected
to be of the same order as the point in time in which the description
given in Sec.(\ref{sub:Short-Time-Scales}) breaks down (see (\ref{eq:kn<1})).
We can therefore obtain an estimation for the order of magnitude of
$t_{hop}$ by inserting Eq.(\ref{eq:Pstat}) into (\ref{eq:kn<1}).
This procedure for the $n^{the}$ resonance, yields $t_{hop}\propto\varepsilon\frac{\sqrt{N}P_{n}^{res}}{k_{n}2A}$.
Using a result from Sec. \ref{sub:Short-Time-Scales} (below Eq.(\ref{eq:EOM_3})),
we can conclude, 
\begin{equation}
t_{hop}=\frac{1}{k_{n}\Delta}\mathcal{O}\left(\frac{\varepsilon}{\eta}\right).\label{eq:thop}
\end{equation}
Equation \ref{eq:thop} is the important relation between parameters
of the system $\left(\eta,k_{n},\Delta\right)$, the initial conditions
(which determine $\varepsilon$) and the hopping time.

This description is strongly supported by numerical results, as explained
in the following section.

\section{Numerical Study \label{sub:Comparison-to-Numerics}}

In this section we study the dynamics of Eq. (\ref{eq:Full Hamiltonian})
by numerical methods. We randomly choose values of $k_{m},\omega_{m}$
and $\phi_{m}$ in Eq. (\ref{eq:V(x,t)}), and solve the corresponding
equations of motion by direct numerical integration. We repeat this
procedure over many realizations of the random parameters and over
different initial conditions. 

We focus on uniform distribution of $k_{m},\omega_{m}$ and $\phi_{m}$,
such that $k_{m}\in\left[-k_{0},k_{0}\right]$, $\omega_{m}\in\left[-\omega_{0},\omega_{0}\right]$
and $\phi_{m}\in\left[0,2\pi\right)$. We take the number of terms
in \eqref{eq:V(x,t)} to be $N\leq10$, in order to ensure that we
are far from the continuum limit. We choose the parameters $k_{0},\omega_{0}$
and $A$ to satisfy the condition
\[
\frac{k_{0}}{\omega_{0}}\sqrt{\frac{8A}{\sqrt{N}}}\leq1,
\]
which is consistent with (\ref{eq:delta}). We have repeated this
procedure for a large variety of parameters. The results that are
presented here were obtained for the parameter set $A=10,k_{0}=\omega_{0}=0.1,N=10$.

\begin{figure}[h]
\begin{centering}
\includegraphics[width=1\columnwidth]{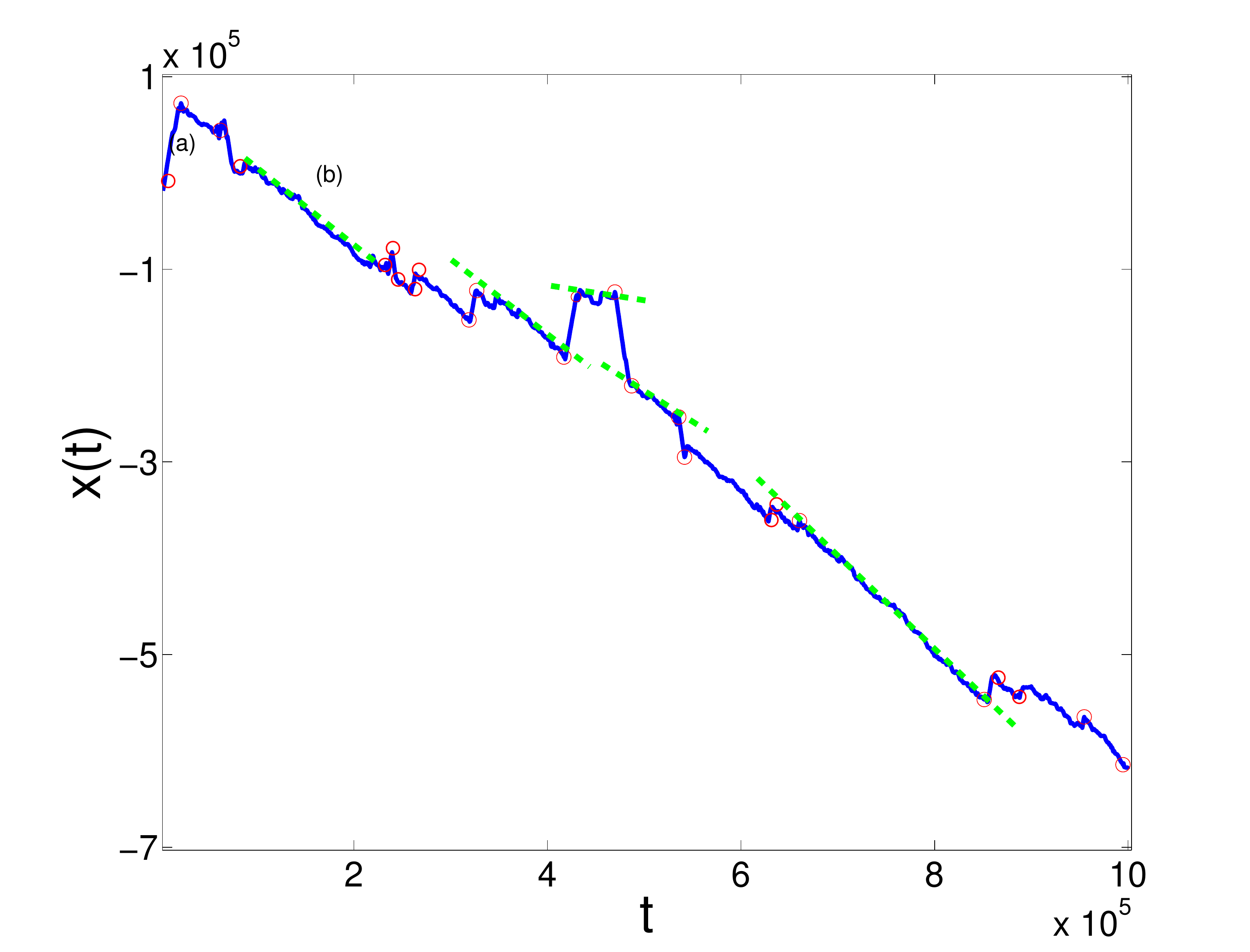}
\par\end{centering}

\centering{}\includegraphics[width=1\columnwidth]{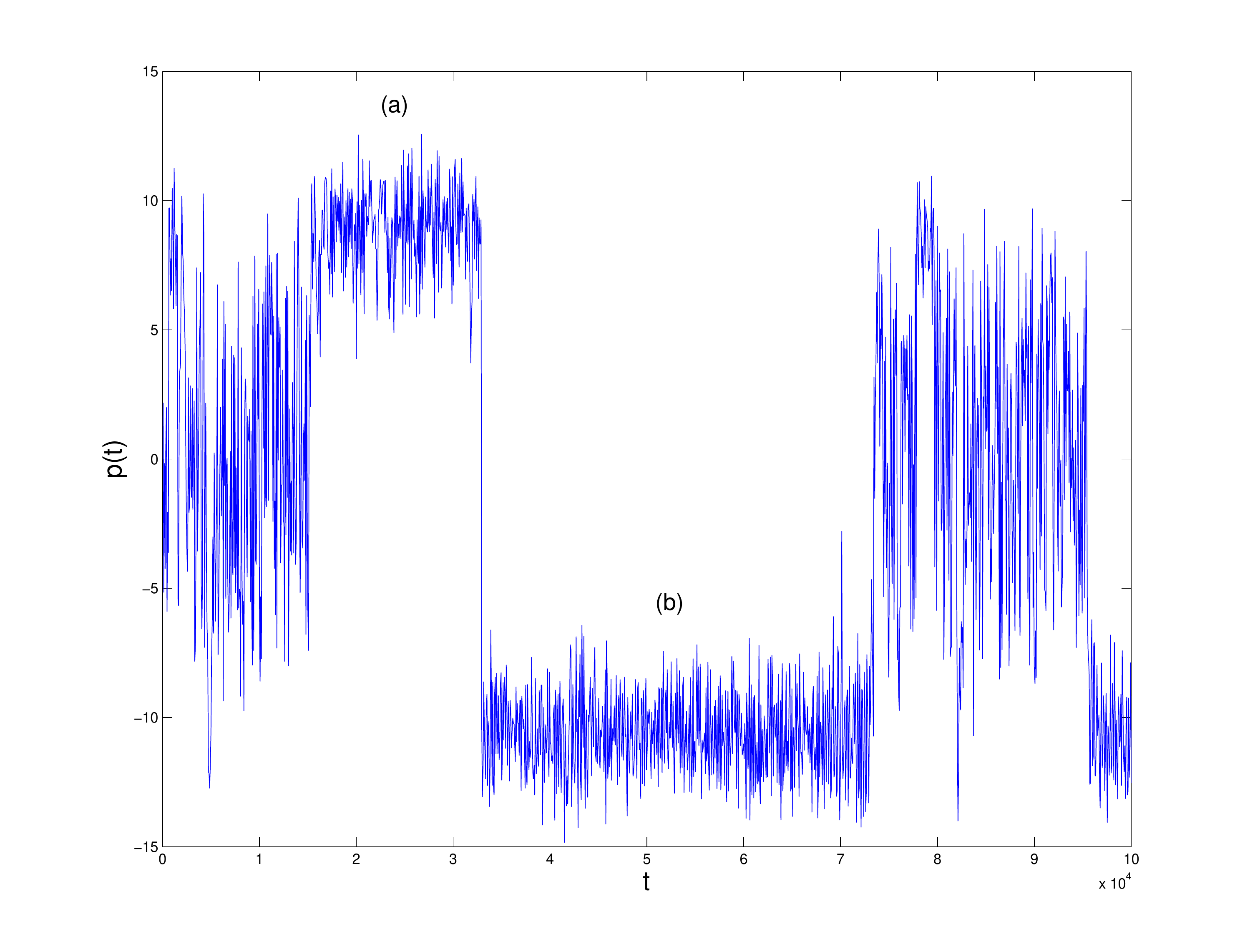}\protect\caption{\label{fig:Two-examples-of}Examples of numerical solutions of the
equations of motion. The upper panel depicts the trajectory as a function
of time. The dashed lines depict the fitted slopes, corresponding
to Eq. (\ref{eq:26}). The lower panel\textbf{ }depicts momentum as
a function of time, over a short time scale. Note that (a) and (b)
denote the same segments in both figures.}
\end{figure}

\begin{figure}[h]
\begin{centering}
\includegraphics[width=1\columnwidth]{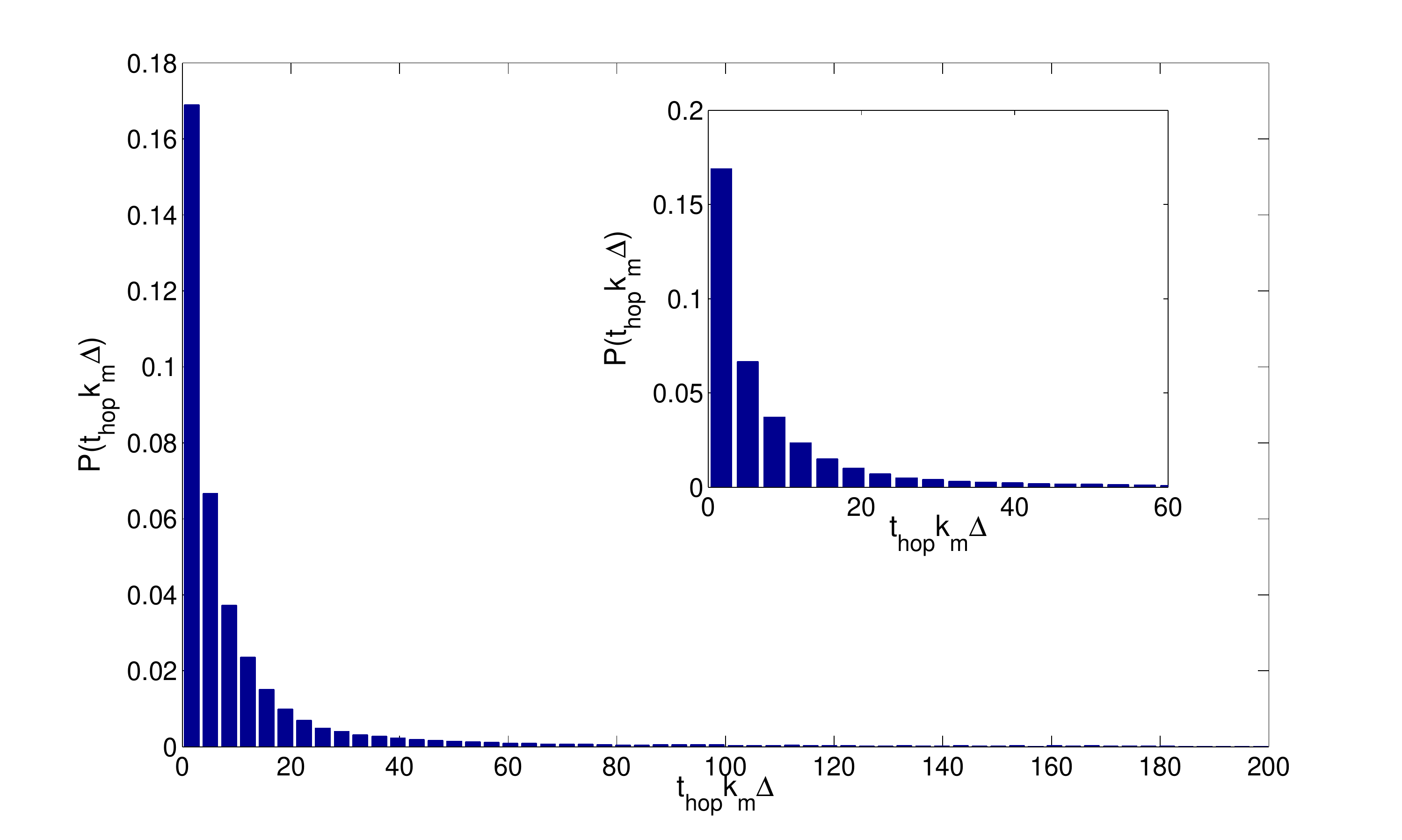}
\par\end{centering}

\protect\caption{\label{fig:Probability-distribution-for t_hop-scaled} Distribution
of the numerical values of the hopping time, $t_{hop}k_{m}\Delta$
(see Eq.(\ref{eq:thop}))}
\end{figure}

Figure \ref{fig:Two-examples-of} depicts an example of a trajectory
in this system. We find that, in agreement with the main result of
Secs. \ref{sub:Short-Time-Scales} and \ref{sub:Finite-Time-Scales},
the trajectories of particles in these setups are composed of linear
segments, in which momentum weakly oscillates around constant values.
The length of each interval where the slope is approximately a constant
corresponds to the previously defined $t_{hop}$ (see Sec. \ref{sub:Finite-Time-Scales},
Eq.(\ref{eq:thop})) . The distribution of the values of $t_{hop}$
near the $m$th resonance, multiplied by $k_{m}\Delta$ is depicted
in Fig. \ref{fig:Probability-distribution-for t_hop-scaled}. We find
that $t_{hop}k_{m}\Delta$ has an average which is typically $\mathcal{O}\left(10^{-1}\right)$,
as shown in Fig. \ref{fig:Probability-distribution-for t_hop-scaled}.
This result suggests that the rate of hops is small, as implied in
Sec. \ref{sub:Finite-Time-Scales}. 

\begin{table}[h]
\begin{tabular}{|c|c|c|}
\hline 
Numerical & Analytic & $P_{n}^{res}$\tabularnewline
\hline 
\hline 
-10.9149 & -10.4243 & -8.4742\tabularnewline
\hline 
-0.8898 & -0.9152 & -0.93063\tabularnewline
\hline 
0.1423 & 0.1251 & 0.1266\tabularnewline
\hline 
-9.4296 & -10.4642 & -8.5484\tabularnewline
\hline 
-0.3535 & -0.3577 & 0.2533\tabularnewline
\hline 
-1.0523 & -1.2855 & -1.5490\tabularnewline
\hline 
-7.0634 & -6.3443 & -5.1400\tabularnewline
\hline 
\end{tabular}

\protect\caption{\label{tab:table_comp}Comparison between the numerical value of the
slopes of the segments which compose the trajectory presented in Fig.\ref{fig:Two-examples-of},
and the analytic estimation (\ref{eq:26}). For comparison, we provide
the value of the corresponding $P_{n}^{res}$ for each segment.}
\end{table}

\begin{figure}[h]
\begin{centering}
\includegraphics[width=1\columnwidth]{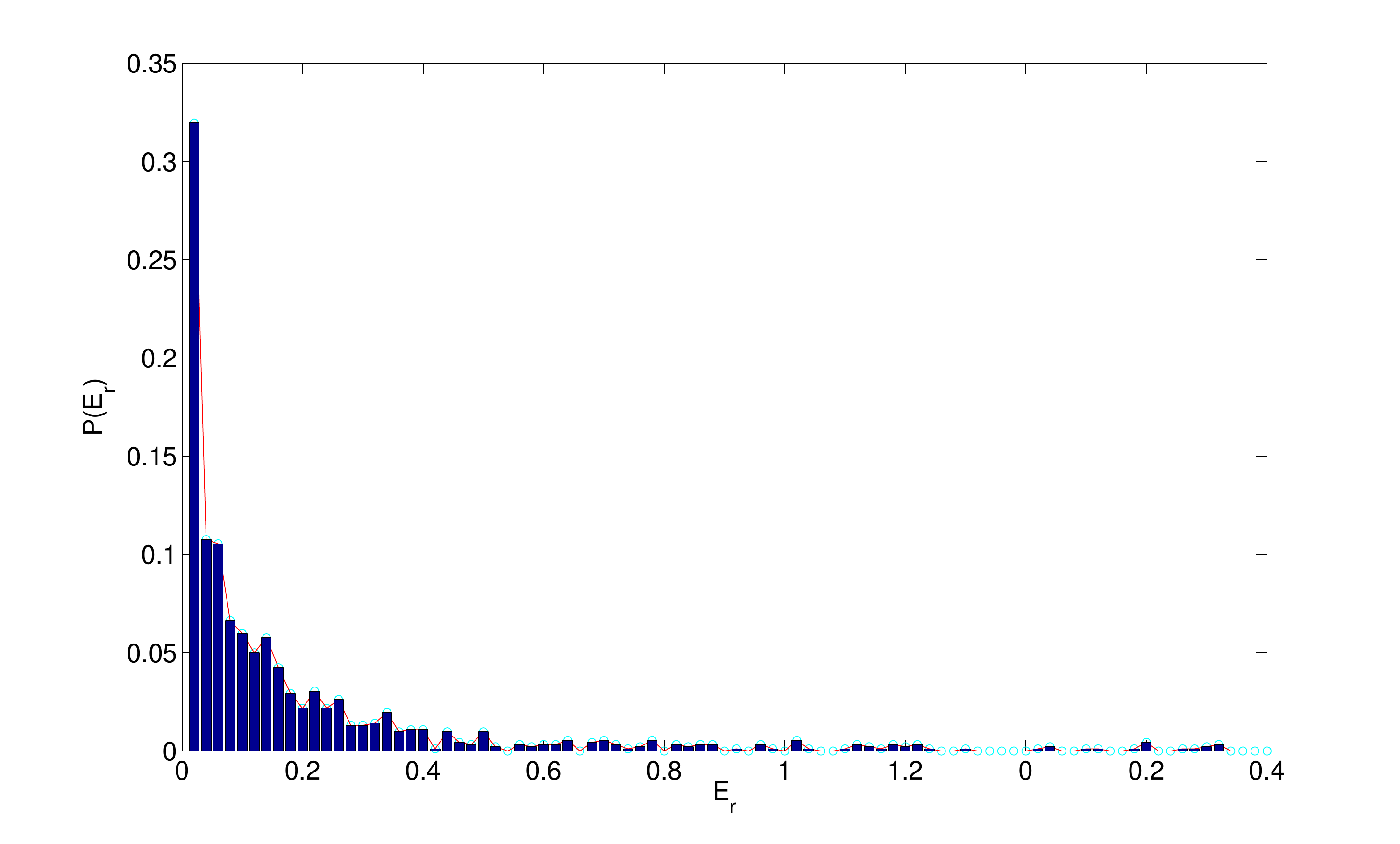}
\par\end{centering}

\protect\caption{The distribution of the deviation between analytical and numerical
results, where $E_{r}$ is given by Eq. (\ref{eq:E_r}). \label{fig:Relative Erro}}
\end{figure}

In order to validate the main result of the analytic calculation,
we decompose each trajectory of the numerical solution into linear
segments as shown in Fig. \ref{fig:Two-examples-of}, by using piece-wise
linear spline methods \cite{ertel1976some}. We then fit each segment
to a linear function and compare the result to Eq. (\ref{eq:26}).
We find a good agreement between Eq. (\ref{eq:26}) and the slopes
of the linear segments which compose the trajectory, see for example
Fig. \ref{fig:Two-examples-of} and table $\text{I}$. We repeat this
procedure over many realizations and calculate the relative error
between the numerical results and Eq. (\ref{eq:26}), 
\begin{equation}
E_{r}=\left|\frac{p_{stat}-p_{num}}{p_{stat}}\right|.\label{eq:E_r}
\end{equation}
The distribution of $E_{r}$ is presented in Fig.\ref{fig:Relative Erro}.

\begin{figure}[h]
\includegraphics[width=1\columnwidth]{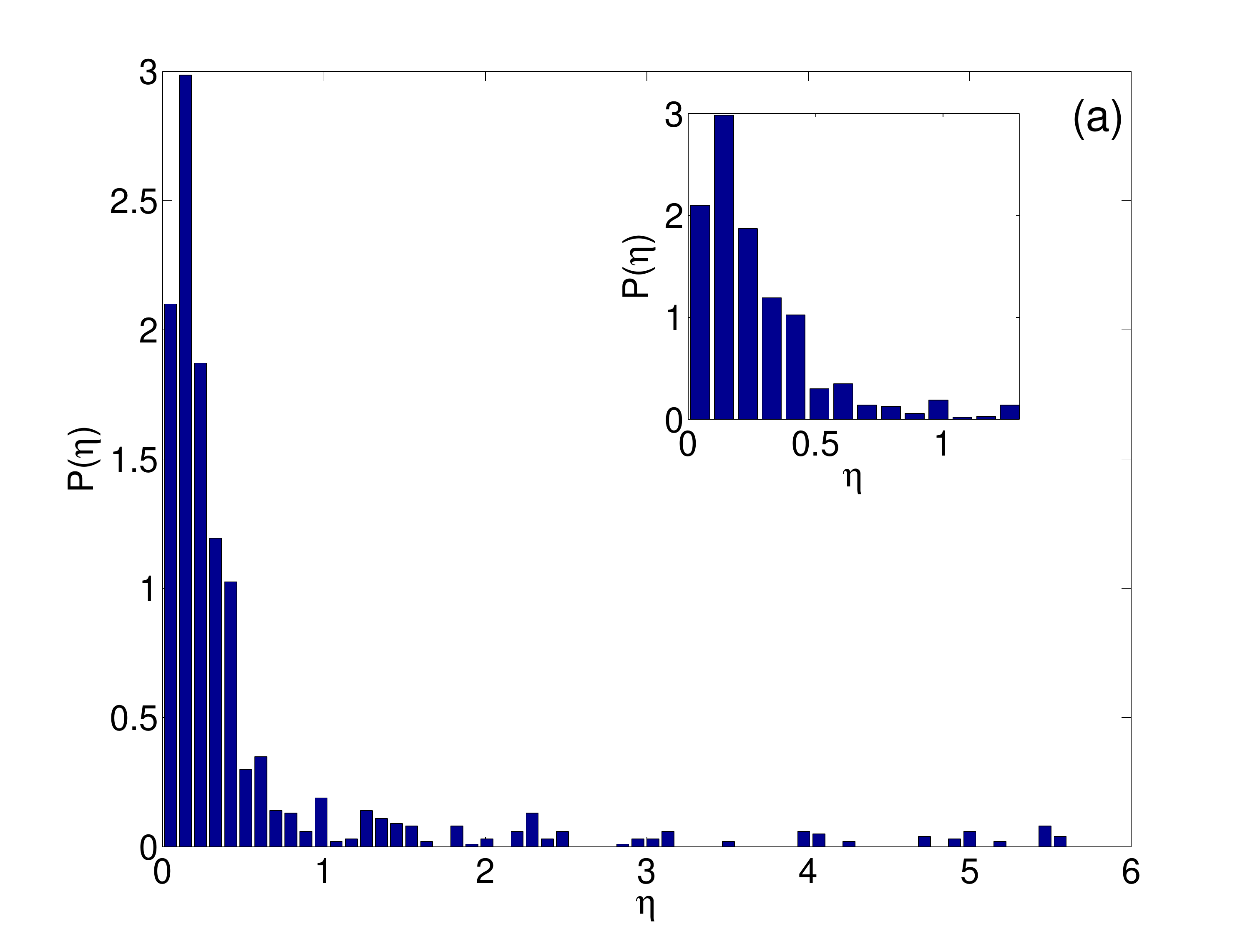}

\includegraphics[width=1\columnwidth]{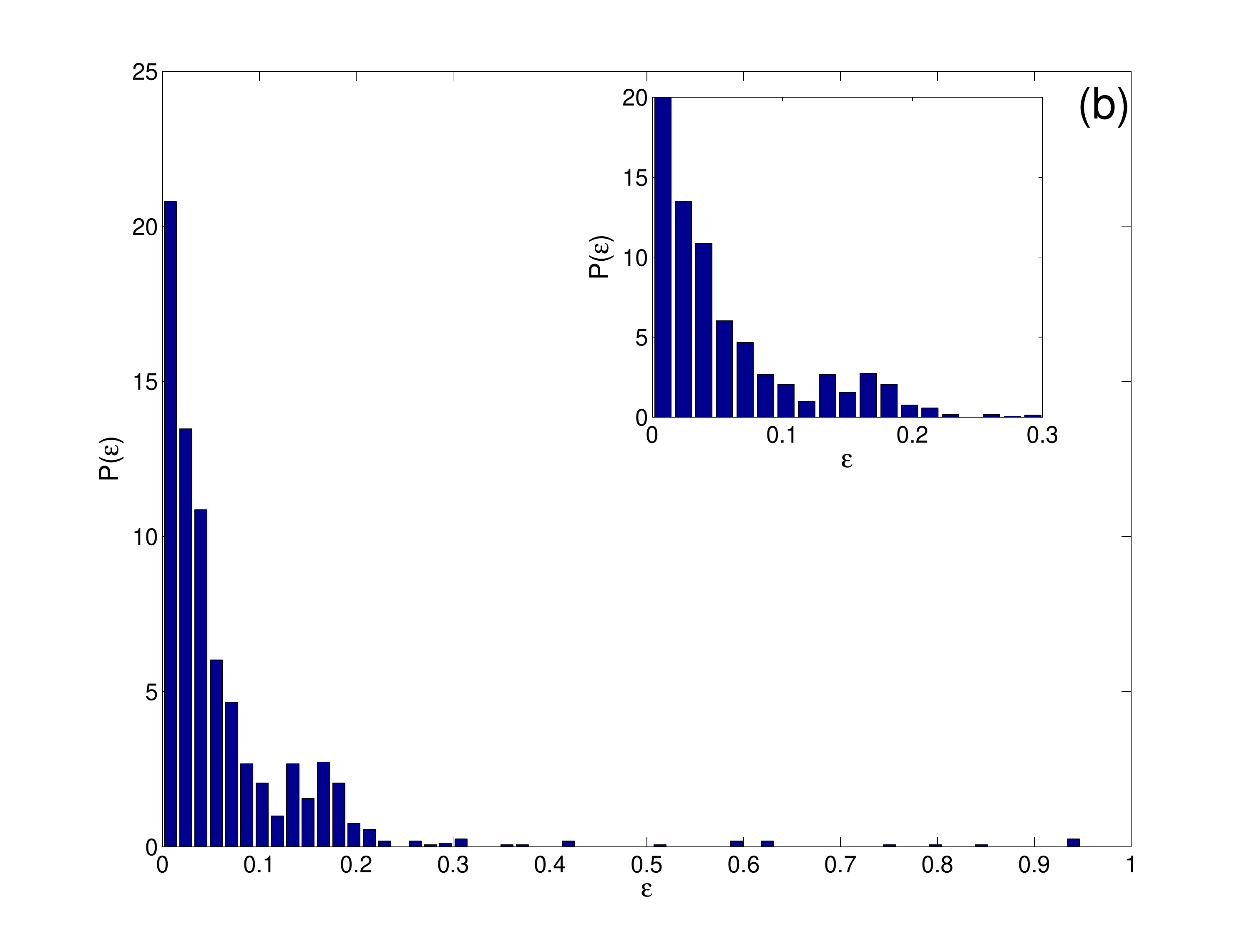}

\protect\caption{The distributions of the small parameters for the studied systems.
\textbf{(a)} The distribution of $\eta$ \textbf{(b)} The distribution
of $\varepsilon$. The insets depict the zooms on smaller regimes.\label{fig:P_delta}}
\end{figure}

We turn to validate the assumptions of this work. The analysis presented
in Sec. \ref{sub:Finite-Time-Scales} relies on the assumption that
the resonances are separated, (see \eqref{eq:delta-1}). We therefore
calculate $\eta$ for different resonances for many different realizations.
We find that for the studied parameter space, the distribution function
of $\eta$ is concentrated around $0.05$, as shown in Fig.\ref{fig:P_delta}.
We can therefore conclude that (\ref{eq:delta-1}) is satisfied for
the parameter space considered in our work.

\begin{figure}[h]
\begin{centering}
\includegraphics[width=0.9\columnwidth]{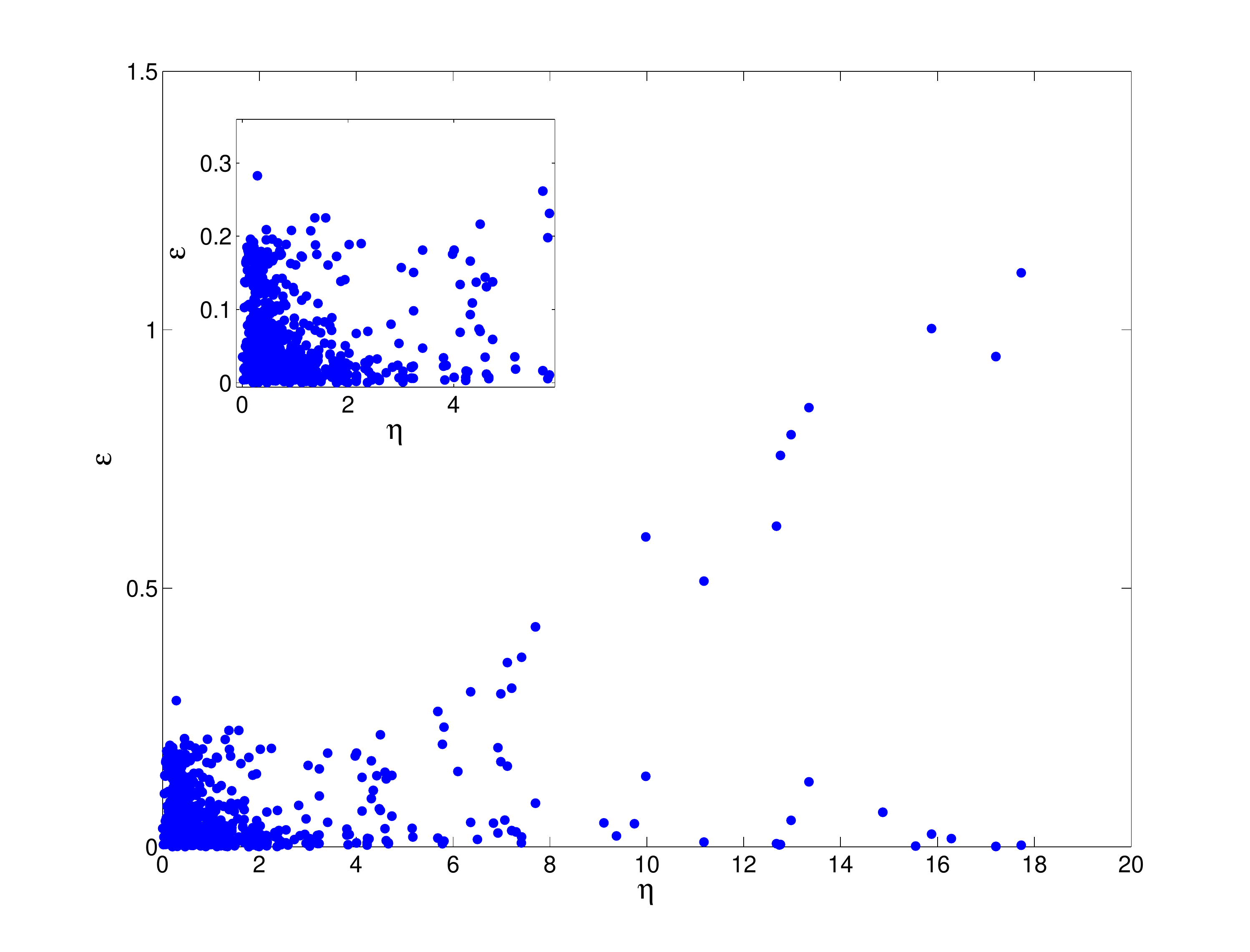}
\par\end{centering}

\protect\caption{\label{fig:scatter_eta_epsilon} Scatter plot of $\varepsilon$ Vs
$\eta$}
\end{figure}

\begin{figure}[h]
\begin{centering}
\includegraphics[width=0.9\columnwidth]{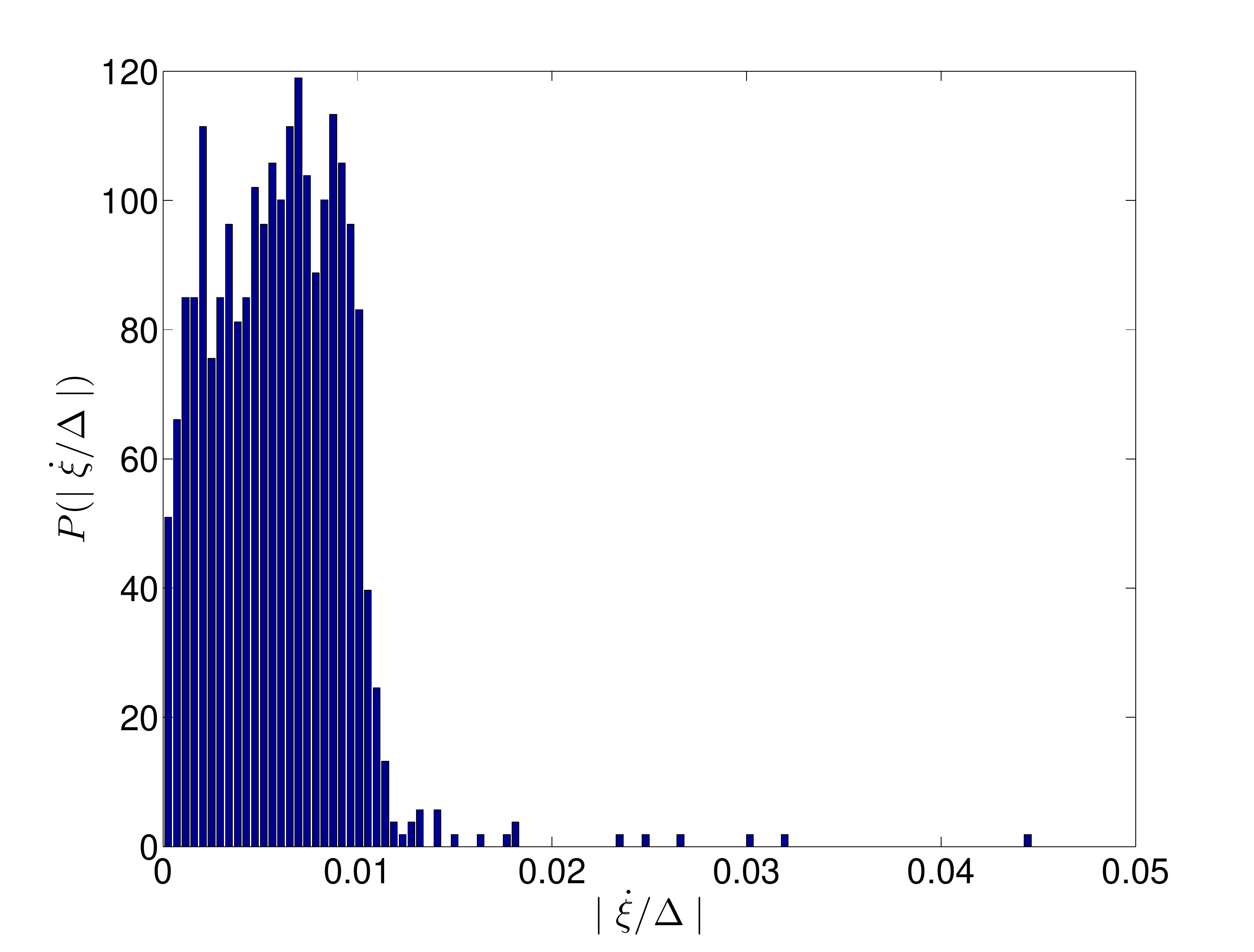}
\par\end{centering}

\protect\caption{\label{fig:Probability-distribution-for xi} Distribution of the oscillatory
components of momentum, normalized by the characteristic resonance
width, $\left|\dot{\xi}/\Delta\right|$. }
\end{figure}

Figure \ref{fig:P_delta}(b) depicts the distribution of $\varepsilon$,
the second small parameter in this work, evaluated over many different
linear segments. We find that the distribution of $\varepsilon$ is
strongly localized near $0$, with a standard deviation which is typically
$\mathcal{O}\left(10^{-2}\right)$. This result is consistent with
(\ref{eq:Eps}) and supports the assumption by which $\varepsilon$
is a small number. Figure \ref{fig:scatter_eta_epsilon} depicts a
scatter of $\varepsilon$ and $\eta$, which allows a comparison between
the relative magnitudes of the small parameter. The scatter of the
values of $\eta$ and $\epsilon$ is presented in Fig. \ref{fig:scatter_eta_epsilon}.

In Sec.\ref{sub:Short-Time-Scales}, we have argued that the oscillatory
component of the momentum, $\dot{\xi}$ can be taken self consistently
to satisfy $\frac{\dot{\xi}}{\Delta}=\mathcal{O}\left(\varepsilon\right)$
and $k_{n}\xi=\mathcal{O}\left(\varepsilon\right)$. We examine this
assumption numerically; for each segment, we subtract the calculated
slope of the segment and examine the remaining oscillatory component.
Fig.\ref{fig:Probability-distribution-for xi} depicts the probability
density function of $\left|\frac{\mbox{\ensuremath{\dot{\xi}}}}{\Delta}\right|$.
We find that indeed, the distribution of $\left|\frac{\mbox{\ensuremath{\dot{\xi}}}}{\Delta}\right|$
is strongly localized around an average of $\mathcal{O}\left(10^{-2}\right)$,
and is therefore commonly a small parameter. A similar result is found
for $k_{n}\xi$.

\section{Summary and Discussion \label{sec:Summary-and-Discussion}}

In this we work we calculated the trajectories of classical particles
under the action of the potential (\ref{eq:V(x,t)}), by estimating
the momentum of particles near Chirikov resonances . We found that
for short time scales the momentum satisfies Eq. (\ref{eq:26}), provided
the conditions (\ref{eq:delta}), (\ref{eq:Eps}) and (\ref{eq:delta-1})
are satisfied. For these time intervals the position is such that
$\xi\left(t\right)$ is of order $\varepsilon$, satisfying (\ref{eq:Eps}).
If the initial conditions are such that the momentum is near a resonance
and $k_{n}\chi\left(0\right)$ is of order $\varepsilon$ the trajectory
will remain near a resonance for a time interval of the order $t_{hop}$,
otherwise it moves chaotically until it approaches the vicinity of
a region in which these conditions are satisfied. It is the main result
of this work. This was verified numerically, and in particular, it
was demonstrated that there is a wide range of parameters for which
Eq. (\ref{eq:26}) holds. We find that on longer time scales, hopping
between Chirikov resonances takes place.

It is interesting to compare our results to previous studies, which
examined Eq.(\ref{eq:V(x,t)}) in two different limits: extremely
small amplitudes and infinite number of overlapping resonances. 

In the limit of extremely small amplitudes, $A_{m}\rightarrow0,\,\forall m$
in Eq.(\ref{eq:V(x,t)}), it was predicted \cite{zaslavskiui1972stochastic,Chirikov1979263}
that the momentum of a particle will remain localized in phase space.
This outcome can be derived from our random walk model by taking the
limit of an extremely weak potential (\ref{eq:V(x,t)}), $A\rightarrow0$
and correspondingly, $\eta\rightarrow0$. In this scenario, no overlap
between resonances exists and the hopping is suppressed. As a result,
the random walk is replaced with localization in phase space such
that $p\left(t\right)=p_{stat}+\mathcal{O}\left(\varepsilon\right)$
remains localized around a single resonance in phase space. This result
is reflected in Eq.(\ref{eq:thop}), as in the limit $\eta\rightarrow0$
and fixed $\varepsilon$, one finds that $t_{hop}\rightarrow\infty$.

Earlier work \cite{krivolapov2012transport,krivolapov2012universality}
focuses on the behavior of for long time and large $N$. In this limit
the number of resonances becomes infinite, and the motion of the particle
was found to obey anomalous diffusion in phase space. Since in the
large $N$ limit, the resonances become dense in phase space and the
rate of hopping between resonances becomes rapid. In this case, the
high-rate random walk results in diffusion in phase space \cite{Diffusion_and_Reactions_in_Fractals}.

The result of the present work can be a starting point of the analysis
of the case where there are few weakly overlapping resonances, which
is opposite to the one studied in previous work \cite{levi-naturephys-2012}.
This is a mixed system, where the motion in some parts of the phase
space is regular, while in other parts it is chaotic \cite{tabor,book_mechlichtenberglieberman,RevModPhys.64.795}.
The Chirikov theory \cite{Chirikov1979263,zaslavskiui1972stochastic}
is not applicable for this case. In addition, the Poincaré-Birkhoff
scenario for generation of chaos \cite{tabor} is not applicable here,
since this system is time dependent with incommensurate periods. 

The results of this work provide a more complete picture of the dynamics
in potentials of the form (\ref{eq:V(x,t)-1}). 

We thank Yevgeny Krivolapov (Bar-Lev) for illuminating communications.
This work was partly supported by the Israel Science Foundation (ISF
- 1028), by the US-Israel Binational Science Foundation (BSF -2010132),
by the USA National Science Foundation (NSF DMS 1201394)and by the
Shlomo Kaplansky academic chair.

\bibliographystyle{unsrt}
\bibliography{EPJB_ReSubmission}

\end{document}